\documentclass[12pt]{article}
\usepackage{subfloat}
\pdfoutput = 1
\usepackage{graphics}
\usepackage{graphicx} 
\usepackage{multirow}
\usepackage{hhline}
\begin{document}
\date{Today}
\title{{\bf{\Large  Noncommutative effects of charged black hole on holographic superconductors }}}

\author{ {\bf {\normalsize Diganta Parai}$^{a}$
\thanks{digantaparai007@gmail.com}},\,
{\bf {\normalsize Debabrata Ghorai}$^{b}$
\thanks{debanuphy123@gmail.com, debabrataghorai@bose.res.in}},\,
{\bf {\normalsize Sunandan Gangopadhyay}$^{b}
$\thanks{sunandan.gangopadhyay@gmail.com, sunandan.gangopadhyay@bose.res.in}}\\
$^{a}$ {\normalsize Indian Institute of Science Education and Research Kolkata}\\{\normalsize Mohanpur, Nadia 741246, India}\\[0.2cm] 
$^{b}$ {\normalsize  S.N. Bose National Centre for Basic Sciences,}\\{\normalsize JD Block, 
Sector III, Salt Lake, Kolkata 700106, India}\\[0.2cm]
}
\date{}

\maketitle

\begin{abstract}
{\noindent In this paper, we analytically investigate the noncommutative effects of a charged black hole on holographic superconductors. The effects of charge of the black hole is investigated in our study. Employing the Sturm-Liouville eigenvalue method, the relation between the critical temperature and charge density is analytically investigated. The condensation operator is then computed. It is observed that condensate gets harder to form for large values of charge of the black hole.}
\end{abstract}
\vskip 1cm

\section{Introduction}
In the past two decades, a lot of investigation has been carried out on the AdS/CFT correspondence \cite{adscft1}-\cite{adscft4} which is a duality between a gravity theory and a gauge theory living on its boundary. In the course of time, it has proved to have tremendous applications in condensed matter physics and also other branches of theoretical physics \cite{hs6}-\cite{dg5}. It is a map which relates strongly coupled systems to weakly coupled systems. The usefulness of this map is evident from the fact that studying the properties of a weakly coupled system can in principle yield the properties of a strongly coupled system. This theoretical insight has been exploited to explain the properties of high $T_{c}$ superconductors which cannot be explained by the BCS theory of superconductivity (known to explain weakly coupled superconductors) \cite{bcs}. The idea is to construct a gravity theory in one higher dimension and study its properties. The map is then applied to extract the properties of the boundary theory. The gravity theory being a weakly coupled system can be investigated by perturbation theory which in turn gives the properties of the strongly coupled system living on the boundary of the gravity theory.
In the gravitational side, a phase transition via scalar hair formation occurs outside the black hole which is asymptotically $AdS$ through the spontaneous symmetry breaking mechanism \cite{hs1}, \cite{hs2}. A lot of investigations has since then been carried out numerically and analytically in this direction \cite{hs6}-\cite{hs5}, \cite{hs16}-\cite{dg6} trying to explain the properties of high $T_{c}$ superconductors. However, one must accept the fact that this model is too crude to make any detailed comparison with any real world material. 

\noindent Another prominent area of research in theoretical physics is noncommutative (NC) geometry, first formally introduced in \cite{ncex1} back in 1947. However, it remained unnoticed till the end of the last century when it was resurrected by giving the rules to move from ordinary quantum field theory (QFT) to NC QFT \cite{ncex2}. In more recent times, a NC inspired Schwarzschild and Reissner-Nordstr\"{o}m (RN) black hole spacetimes were obtained in \cite{nico1,nico2}. Here the Einstein's equation of general relativity were solved by incorporting the effect of noncommutativity through a smeared matter source. An interesting feature of this black hole solution is that it does not have a physical singularity. In \cite{rb1}, an in depth study of the thermodynamics of this black hole solution was made.

\noindent In this paper, we have analytically investigated the effect of noncommutativity of a NC inspired charged black hole on holographic superconductors. The motivation of this work is the following. We want to study whether the charge of a black hole favours the formation of the condensate. The effects of noncommutativity of a NC inspired Schwarzshild black hole on holographic superconductors have been investigated in \cite{dg2, subir}. In this paper, the analysis has been extended to the case of charged black holes. We also carry out our study in the framework of Born-Infeld (BI) electrodynamics. The calculations are valid upto first order in the BI parameter. Employing the Sturm-Liouville eigenvalue method, we obtain the relation between the critical temperature and the charge density. We observe that the condensation gets hard to form in the presence of the BI parameter $b$, NC parameter $\theta$ and mass $M$ and charge $Q$ of the black hole. We also find that there exists an upper bound of the charge of the black hole above which we do not get a physical solution. Finally, we compare our analytical results with the numerical results.

\noindent The paper is organized as follows. In section 2, we provide the basic holographic set up for the holographic superconductors in the presence of NC inspired Reissner-Nordstrom black hole. In section 3, we compute the critical temperature in terms of a solution to the Sturm-Liouville eigenvalue problem. In section 4, we analytically obtain an expression for the condensation operator. Finally, we summarize our findings in section 5.



\section{Basic set up in noncommutative space  }
\noindent The action for the formation of scalar
hair in an electrically charged black hole in $d$-dimensional anti-de Sitter spacetime in the framework of Born-Infeld electrodynamics reads 
\begin{eqnarray}
S=\int d^{d}x \sqrt{-g}\left[\frac{(R+2\Lambda)}{16\pi G_{d}} + \frac{1}{b}\left(1-\sqrt{1+ \frac{b}{2}F^{\mu \nu} F_{\mu \nu}}\right) - (D_{\mu}\psi)^{*} D^{\mu}\psi-m^2 \psi^{*}\psi \right]
\label{dp1}
\end{eqnarray}
where $\Lambda=(d-1)(d-2)/(2L^2)$ is the cosmological constant, $F_{\mu \nu}=\partial_{\mu}A_{\nu}-\partial_{\nu}A_{\mu}$ is the field strength tensor, 
$D_{\mu}\psi=\partial_{\mu}\psi-iA_{\mu}\psi$ is the covariant derivative,  $A_{\mu}$ and $ \psi $ represent the gauge and the scalar fields.\\
\noindent The black hole spacetime in which we consider the above action is that of a $d$-dimensional planar NC  RN $AdS_{d}$ black hole whose metric reads \cite{nico1}, \cite{nico2}
\begin{eqnarray}
ds^2=-f(r)dt^2+\frac{1}{f(r)}dr^2+ r^2  dx_{i} dx^{i}\nonumber
\end{eqnarray}
\begin{eqnarray}
f(r)=\frac{r^2}{L^2}-\frac{2MG_{d}}{r^{d-3}}\frac{\gamma \left(\frac{d-1}{2},\frac{r^2}{4\theta }\right)}{\Gamma( \frac{d-1}{2})}+\frac{(d-2)(d-3)^2G_{d}Q^{2}}{2\pi^{d-3}r^{2(d-3)}}\left[F(r)+E_{d}r^{d-3}\gamma \left(\frac{d-1}{2},\frac{r^2}{4\theta }\right)\right]
\label{dp2}
\end{eqnarray}
where $dx_{i}dx^{i}$ represents the line element of $(d-2)$-dimensional hypersurface with no curvature and 
\begin{eqnarray}
\label{dp3}
E_{d}=\frac{2^{\frac{11-3d}{2}}}{(d-3)\theta ^{\frac{d-3}{2}}}\frac{\Gamma(\frac{d-3}{2})}{\Gamma (\frac{d-1}{2})}~&;&~
F(r)=\gamma ^{2}\left(\frac{d-3}{2},\frac{r^2}{4\theta }\right)-\frac{2^{\frac{11-3d}{2}}r^{d-3}}{(d-3)\theta ^{\frac{d-3}{2}}}\gamma \left(\frac{d-3}{2},\frac{r^2}{2\theta }\right)\\
\gamma(s,x)&=&\int_{0}^{x}t^{s-1}e^{-t}dt
\label{dp4}
\end{eqnarray}  
is the lower incomplete Gamma function, $M$ is the mass of the black hole and $Q$ the charge of the black hole. In the limit $\theta\to 0$, one recovers the commutative RN black hole spacetime in $d$ dimensions.\\ 
\noindent The Hawking temperature, interpreted as the temperature of the conformal field theory on the boundary, is given by
\begin{eqnarray}
T_{H}=\frac{f^{\prime}(r_{+})}{4\pi}
\label{dp5}
\end{eqnarray}
where $r_{+}$ is the event horizon radius of the black hole.\\
\noindent  In the rest of our calculation, we shall set $L=1$ for convenience. The radius of the event horizon of the black hole can be obtained from the relation $f(r_+)=0$ and reads
\begin{eqnarray}
r_{+}^{d-1} = \gamma \left(\frac{d-1}{2},\frac{r^{2}_{+}}{4\theta}\right)\left[\frac{2MG_d}{\Gamma(\frac{d-1}{2})} - \frac{(d-2)(d-3)^2 }{2 \pi^{d-3}}G_d Q^2 E_d \right] &-& \frac{(d-2)(d-3)^2 }{2 \pi^{d-3}} \frac{F(r_+}{r_{+}^{(d-3)}} G_d Q^2 \nonumber \\
\Rightarrow \gamma \left(\frac{d-1}{2},\frac{r^{2}_{+}}{4\theta}\right) = \frac{r_{+}^{d-1}\left[1+ \frac{(d-2)(d-3)^2 }{2 \pi^{d-3}} \frac{F(r_+}{r_{+}^{2(d-2)}} G_d Q^2\right]}{\frac{2MG_d}{\Gamma(\frac{d-1}{2})}\left[1- \frac{(d-2)\Gamma(\frac{d-1}{2})}{(8\theta\pi^2)^{\frac{d-3}{2}}}\frac{Q^2}{M}\right]}~.&&
\end{eqnarray}
Using this relation and eq.(\ref{dp5}), we get the expression for the Hawking temperature of the black hole to be
\begin{eqnarray}
T_{H}=\frac{r_{+}}{4\pi}\Bigg[(d-1)-\frac{4MG_{d}}{\Gamma (\frac{d-1}{2})}\frac{e^{-{\frac{r_{+}^2}{4\theta}}}}{(4\theta)^{\frac{d-1}{2}}}\bigg\{1-\frac{(d-2)(d-3)\Gamma (\frac{d-3}{2})}{2^{\frac{3d-7}{2}}\pi^{d-3}\theta^{\frac{d-3}{2}}}\frac{Q^2}{M}\bigg\}\nonumber\\
-\frac{2(d-2)(d-3)Q^{2}G_{d}}{\pi^{d-3}r_{+}^{2(d-2)}}\gamma ^{2}\bigg(\frac{d-1}{2},\frac{r_{+}^2}{4\theta}\bigg)\Bigg].
\label{dp6}
\end{eqnarray}

\noindent The ansatz for the gauge field and the scalar field is now chosen to be \cite{hs6}
\begin{eqnarray}
A_{\mu} = (\phi(r), 0, 0, 0)~,~~~\psi=\psi(r).
\label{dp7}
\end{eqnarray}
The equations of motion for the gauge and the scalar fields with this ansatz read
\begin{eqnarray}
\phi^{\prime \prime}(r) + \frac{d-2}{r}\phi^{\prime}(r) - \frac{d-2}{r}b\phi^{\prime}(r)^{3} - \frac{2  \phi(r) \psi^{2}(r)}{f(r)}(1 - b  \phi^{\prime}(r)^{2})^\frac{3}{2} = 0
\label{dp8}
\end{eqnarray}
\begin{eqnarray}
\psi^{\prime \prime}(r) + \left(\frac{d-2}{r}+ \frac{f^{\prime}(r)}{f(r)}\right)\psi^{\prime}(r) + \left(\frac{\phi^{2}(r)}{f(r)^2}- \frac{m^{2}}{f(r)}\right)\psi(r) = 0
\label{dp9}
\end{eqnarray}
where prime denotes derivative with respect to $r$.\\ 
\noindent The  $\phi(r_+)=0$ and $\psi(r_{+})$ to be finite at the horizon are required for regularization. Near the boundary, $r$ is large and hence $e^\frac{-r^2}{4\theta}\ll 1$ as $\theta$ is small which in turn implies that the effect of noncommutativity of space can be neglect on the asymptotic behaviour of the fields. The asymptotic behaviour of the fields can therefore be written as 
\begin{eqnarray}
\label{dp10}
\phi(r)&=&\mu-\frac{\rho}{r^{d-3}}\\
\psi(r)&=&\frac{\psi_{-}}{r^{\Delta_{-}}}+\frac{\psi_{+}}{r^{\Delta_{+}}}
\label{dp11}
\end{eqnarray}
where
\begin{eqnarray}
\Delta_{\pm}&=&\frac{(d-1)\pm\sqrt{(d-1)^2+4m^2 }}{2}
\label{d12}
\end{eqnarray}
\noindent and $\mu$ and $\rho$ are interpreted as the charge density and the chemical potential of the boundary field theory from the gauge/gravity dictionary. In this paper, we choose $\psi_{-}=0$ so that $\psi_{+}$ is dual to the expectation value of the condensation operator $J$ in the absence of source term for condensation.\\
\noindent Under the change of coordinates $z=\frac{r_{+}}{r}$,  the field equations (\ref{dp8}),(\ref{dp9}) become
\begin{eqnarray}
\phi^{\prime \prime}(z) - \left(\frac{d-4}{z}\right) \phi^{\prime}(z) + \frac{d-2}{r^2_{+}} b \phi^{\prime}(z)^{3} z^3 - \frac{2r^2_{+} \phi(z) \psi^{2}(z)}{f(z) z^4}\left(1 -\frac{b z^4}{r^2_{+}} \phi^{\prime}(z)^{2}\right)^\frac{3}{2}= 0
\label{dp13}
\end{eqnarray}

\begin{eqnarray}
\psi^{\prime \prime}(z) + \left(\frac{f^{\prime}(z)}{f(z)} - \frac{d-4}{z}\right)\psi^{\prime}(z) + \frac{r^2_{+}}{z^4} \left(\frac{\phi^{2}(z)}{f(z)^2}- \frac{m^{2}}{f(z)}\right)\psi(z)=0 
\label{dp14}
\end{eqnarray}
where prime now denotes derivative with respect to $z$. Our aim is to solve these equations in the interval $[0, 1]$, where $z=1$ is the horizon and $z=0$ is the boundary.
Further, we choose $m^{2}=-3$ consistent with the Breitenlohner-Freedman bound \cite{ncex9}. This leads to $\Delta_{+}=3$ and $\Delta_{-}=1$ for $d=5$.



\section{The critical temperature $T_{c}$}
In this section we analyze the relation between the critical temperature and the charge density. To start the analysis, we note that the matter field must vanish for $T\geq T_{c}$. The aim is to investigate the behaviour of $\psi(r)$ just below the critical temperature $T_{c}$. To do that, we first solve the equation for $ \phi(r)$ at $T=T_{c}$ where $\psi(r)$ vanishes. Eq.(\ref{dp13}) therefore reduces to
\begin{eqnarray}
\label{dp15}
\phi^{\prime \prime}(z) -\frac{d-4}{z}\phi^{\prime}(z) + \frac{(d-2)bz^3}{r^2_{+(c)}} \phi^{\prime}(z)^3=0.
\end{eqnarray}
To solve the above equation, we set $\phi^{\prime}(z)=\xi(z)$ and obtain a first order differential equation $\xi(z)$. This reads
\begin{eqnarray}
\label{dp16}
\xi^{ \prime}(z) -\frac{d-4}{z}\xi(z) + \frac{(d-2)bz^3}{r^2_{+(c)}} \xi(z)^3=0.
\end{eqnarray}
This is a Bernoulli differential equation in $\xi(z)$ which can be recast as 
\begin{eqnarray}
\frac{d}{dz}\left(\frac{z^{2(d-4)}}{\xi(z)^2}\right)=\frac{2(d-2)b}{r_{+(c)}^{2}}z^{2d-5} ~.
\label{dp17}
\end{eqnarray}
Integrating the above equation in the interval $[0,1]$ and using the boundary condition (\ref{dp10}) which in terms of $\xi$ takes the form $\xi(0)=-\frac{(d-3)\rho}{r_{+}^{d-3}}$, we obtain
\begin{eqnarray}
\frac{1}{\xi(1)^{2}}=\frac{b}{r_{+(c)}^2}+\left(-\frac{r_{+(c)}^{d-3}}{(d-3)\rho}\right)^2 ~.
\label{dp18}
\end{eqnarray}
Now we integrate eq.(\ref{dp17}) in the interval $[1,z]$ and use eq.(\ref{dp18}) to get
\begin{eqnarray}
\xi(z)=\phi^{\prime}(z)=-\frac{\lambda r_{+(c)}(d-3)z^{d-4}}{\sqrt{1+(d-3)^{2}b \lambda^{2}  z^{2(d-2)}}} 
\label{dp19}
\end{eqnarray}
where 
\begin{eqnarray}
\lambda=\frac{\rho}{r_{+(c)}^{d-2}} ~.
\label{dp20}
\end{eqnarray}
Integrating eq.(\ref{dp19}) once more in the interval $[1,z]$ and using the fact that $\phi(z=1)=0$, we obtain 
\begin{eqnarray}
\phi(z)=-\int_{1}^{z}\frac{\lambda r_{+(c)}(d-3)z^{\prime(d-4)}}{\sqrt{1+(d-3)^{2}b \lambda^{2}  z^{\prime2(d-2)}}} dz^{\prime} ~.
\label{dp21}
\end{eqnarray}
The above integral cannot be done exactly and hence we expand the integrand binomially and keep terms upto first order in the BI parameter $b$. On doing this, the solution for $\phi(z)$ takes the form 
\begin{eqnarray}
\phi(z) &=& \lambda r_{+(c)}\bigg\{(1-z^{d-3}) - \frac{b(\lambda^2|_{b=0}) (d-3)^3}{2(3d-7)} (1-z^{3d-7})\bigg\}
\label{dp22}
\end{eqnarray}
where we have used the fact that $b\lambda^2= b(\lambda^2|_{b=0}) + \mathcal{O}(b^2)$ \cite{sgdc1}, $\lambda^2|_{b=0}$ being the value of $\lambda^2$ for $b=0$. 

\noindent Now we concentrate on the metric. With the change of coordinate $z=\frac{r_{+}}{r}$, the black hole spacetime given by eq.(\ref{dp2}) reads
\begin{eqnarray}
f(z)=\frac{r_{+(c)}^2}{z^2}g_{0}(z)
\label{dp23}
\end{eqnarray}
where
\begin{eqnarray}
g_{0}(z)=&1&-\bigg[\Xi\bigg\{1+\frac{(d-2)(d-3)^{2}G_{d}Q^{2}}{2\pi^{d-3}r_{+(c)}^{2(d-2)}}F(r_{+(c)})\bigg\}+\frac{(d-2)(d-3)G_{d}Q^2}{(8\theta\pi^2)^{\frac{d-3}{2}}r_{+(c)}^{d-1}}\gamma(\frac{d-3}{2},\frac{r_{+(c)}^{2}}{2\theta z^2})\bigg]z^{d-1}\nonumber\\
&+&\frac{(d-2)(d-3)^{2}G_{d}Q^{2}}{2\pi^{d-3}r_{+(c)}^{2(d-2)}}\gamma^2 \bigg(\frac{d-3}{2},\frac{r_{+(c)}^{2}}{4\theta z^2}\bigg)z^{2(d-2)}
\label{dp24}
\end{eqnarray}
\begin{eqnarray}
\Xi=\frac{\gamma \bigg(\frac{d-1}{2},\frac{r_{+(c)}^{2}}{4\theta z^2}\bigg)}{\gamma\bigg(\frac{d-1}{2},\frac{r_{+(c)}^{2}}{4\theta }\bigg)} ~.
\label{dp25}
\end{eqnarray}
Now eq.(\ref{dp14}) for the field $\psi$ near the critical temperature $T\rightarrow T_{c} $ approaches the limit 
\begin{eqnarray}
\psi^{\prime\prime}(z)+\bigg(\frac{g_{0}^{\prime}(z)}{g_{0}(z)}-\frac{d-2}{z}\bigg)\psi^{\prime}(z)+\bigg(\frac{\phi^2(z)}{r_{+(c)}^{2}g_{0}^{2}(z)}-\frac{m^2}{z^{2}g_{0}(z)}\bigg)\psi(z)=0
\label{dp26}
\end{eqnarray}
where $\phi(z)$ now corresponds to the solution in eq.(\ref{dp22}).\\
\noindent Now defining 
\begin{eqnarray}
\psi(z)=\frac{\langle J \rangle}{r_{+}^{\Delta_{+}}}z^{\Delta_{+}}\mathcal{F}(z)
\label{dp27}
\end{eqnarray}
where $\mathcal{F}(0)=1$ and J is the  condensation operator and substituting this form of $\psi(z)$ in eq.(\ref{dp26}), we get
\begin{eqnarray}
\mathcal{F}^{\prime\prime}(z)&+&\Bigg\{\frac{2\Delta_{+}-d+2}{z}+\frac{g_{0}^{\prime}(z)}{g_{0}(z)}\Bigg\}\mathcal{F}^{\prime}(z)+\Bigg\{\frac{\Delta_{+}(\Delta_{+}-1)}{z^2}+\bigg(\frac{g_{0}^{\prime}(z)}{g_{0}(z)}-\frac{d-2}{z}\bigg)\frac{\Delta_{+}}{z}-\frac{m^2}{z^{2}g_{0}(z)}\Bigg\}\mathcal{F}(z)\nonumber\\
&+&\frac{\lambda^{2}\widetilde{\phi}^{2}(z)}{g_{0}^{2}(z)}\mathcal{F}(z)=0
\label{dp28}
\end{eqnarray}
to be solved subject to the boundary condition $\mathcal{F}^{\prime}(0)=0$.\\
\noindent This equation can be rewritten in the Sturm-Liouville form
\begin{eqnarray}
\frac{d}{dz}\big\{p(z)\mathcal{F}^{\prime}(z)\big\}+q(z)\mathcal{F}(z)+\lambda^{2}r(z)\mathcal{F}(z)=0
\label{dp29}
\end{eqnarray}
with
\begin{eqnarray}
&p(z)&=z^{2\Delta_{+}-d+2}g_{0}(z)\nonumber\\
&q(z)&=\Delta_{+}z^{2\Delta_{+}-d}\bigg\{zg_{0}^{\prime}(z)+(\Delta_{+}-d+1)g_{0}(z)-\frac{m^2}{\Delta_{+}}\bigg\}\nonumber\\
&r(z)&=\frac{z^{2\Delta_{+}-d+2}}{g_{0}(z)}\bigg\{(1-z^{d-3})^{2} - \frac{b(\lambda^2|_{b=0}) (d-3)^3}{(3d-7)}(1-z^{d-3}) (1-z^{3d-7})\bigg\} ~.
\label{dp30}
\end{eqnarray}
The task is to estimate the value of $\lambda^2$. The following expression for $\lambda^2$ does the job
\begin{eqnarray}
\lambda^{2}=\frac{\int_{0}^{1}dz\big\{p(z)[\mathcal{F}^{\prime}(z)]^{2}-q(z)[\mathcal{F}(z)]^{2}\big\}}{\int_{0}^{1}dz r(z)[\mathcal{F}(z)]^{2}}
\label{dp31}
\end{eqnarray}
Since extremizing this expression yields eq.(\ref{dp29}). To estimate the eigenvalue $ \lambda^2$, we shall now use the trial function
\begin{eqnarray}
\mathcal{F}=\mathcal{F}_{\alpha}(z)=1-\alpha z^2
\label{dp32}
\end{eqnarray}
with $\mathcal{F}_{\alpha} (z)$ satisfying to $\mathcal{F}_{\alpha}(0)=1$ and $\mathcal{F}_{\alpha}^{\prime}(0)=0$.
 
\begin{figure}[h!]
\centering
\includegraphics[scale=.45]{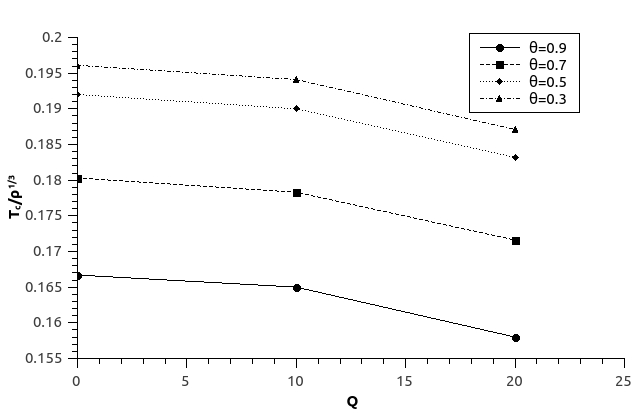}
\includegraphics[scale=.45]{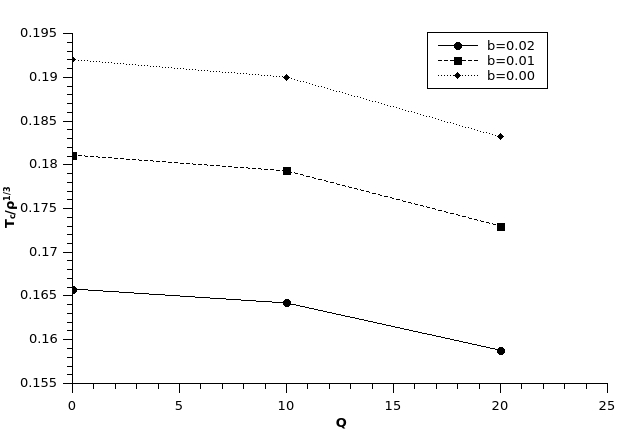}
\includegraphics[scale=.65]{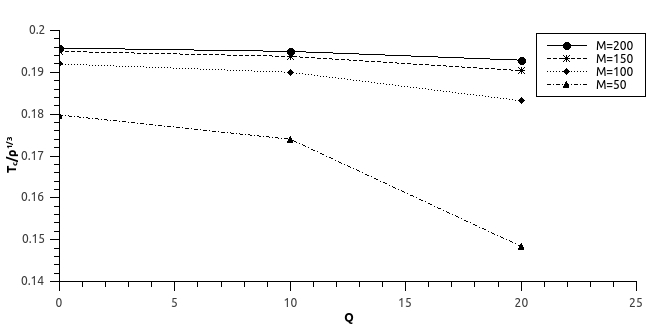}
\caption{The first Figure displays the $\frac{T_c}{\rho^{1/3}}$ vs. $Q$ plot in the framework of Maxwell electrodynamics $(b = 0)$ and $M=\frac{100}{G_5}$ for different values of the NC parameter $\theta$. The second Figure shows the $\frac{T_c}{\rho^{1/3}}$ vs. $Q$ plot with $M=\frac{100}{G_5}$ and $\theta=0.5$ for different values of the BI parameter $b$. In the last Figure, for fixed value of $\theta = 0.5$ and $b=0$, the $\frac{T_c}{\rho^{1/3}}$ vs. $Q$ plot for different values of the mass $M$ of the black hole is displayed. }
\label{dpfig1}
\end{figure}

\noindent We now move onto obtain the relation between the critical temperature and the charge density for $d=5$. At $T=T_{c}$ the horizon radius $r_{+}$ can be obtained from $f(r_{+(c)})=0$. Now since the horizon radius is large compared to the NC length scale, hence we can keep the leading order terms in the Gamma function to obtain 
\begin{eqnarray}
r_{+(c)}^6 &=& 2MG_{5}r_{+(c)}^2\bigg\{1-\bigg(1+\frac{r_{+(c)}^2}{4\theta}\bigg)e^{-\frac{r_{+(c)}^2}{4\theta}}\bigg\} \nonumber \\
&-&\frac{6Q^{2}G_{5}}{\pi^{2}}\bigg[\bigg(1-e^{-\frac{r_{+(c)}^2}{4\theta}}\bigg)^{2}-\frac{r_{+(c)}^2}{8\theta}e^{-\frac{r_{+(c)}^2}{4\theta}}\bigg(1+\frac{r_{+(c)}^2}{4\theta}-e^{-\frac{r_{+(c)}^2}{4\theta}}\bigg)\bigg].
\label{dp33}
\end{eqnarray}\\
In the limit $\theta \rightarrow 0$, the above equation takes the form
\begin{eqnarray}
(r_{+(c)}^{(0)})^{6}=2MG_{5}(r_{+(c)}^{(0)})^2-\frac{6Q^{2}G_{5}}{\pi^{2}}
\label{dp34}
\end{eqnarray}
where $r^{(0)}_{(c)}$ denotes the commutative horizon radius at the critical temperature.
To get a real positive solution for $r^{(0)}_{+(c)}$ we have to put a bound on $Q$ which reads
\begin{eqnarray}
Q\leq \frac{\pi}{\sqrt{3}}\bigg(\frac{2MG_{5}^\frac{1}{3}}{3}\bigg)^\frac{3}{4}~.
\label{dp35}
\end{eqnarray}
The largest root of eq.(\ref{dp34}) with the largest value of $Q$ reads
\begin{eqnarray}
r^{(0)}_{+(c)}=\bigg(\frac{8MG_{5}}{3}\bigg)^\frac{1}{4}\sqrt{\cos\bigg[\frac{1}{3}\bigg(\pi-\tan^{-1} \sqrt{\frac{8\pi^4}{243}\frac{M^{3}G_{5}}{Q^4}-1}\bigg)\bigg]}\equiv R~.
\label{dp36}
\end{eqnarray}
We now consider a small NC correction on the horizon radius of the form
\begin{eqnarray}
r_{+(c)} =R+ \sigma \theta ~.
\label{dp37}
\end{eqnarray}
Substituting this in eq.(\ref{dp33}) and solving for $\sigma$ yields
\begin{eqnarray}
\sigma=\frac{2}{R}\Bigg[\frac{(R^{4}+4R^{2}\theta+64\theta^2)-4\theta(R^2+8\theta)e^{-\frac{R^2}{4\theta}}-\frac{8\pi^2}{3}\frac{MR^{2}\theta}{Q^2}(R^2+4\theta)}{(R^4-4R^{2}\theta+48\theta^{2})-8\theta (R^2+6\theta )e^{-\frac{R^2}{4\theta}}-\frac{8\pi^2}{3}\frac{M\theta}{Q^2}(R^4-4R^{2}\theta-16\theta^2)+\frac{64\pi^{2}}{3G_{5}Q^2}\theta^{3}(3R^{4}-2MG_{5})e^\frac{R^2}{4\theta}}\Bigg].\nonumber \\
\label{dp38}
\end{eqnarray}
Using eq.(s)(\ref{dp36}, \ref{dp37}, \ref{dp20}), we obtain from eq.(\ref{dp6}) upto first order in $\sigma$
\begin{eqnarray}
T_{c}=\frac{1}{\pi}\Bigg[1-\frac{MG_{5}}{(4\theta)^2}e^{-\frac{R^2}{4\theta}}\bigg\{1-\frac{3}{8\pi^2}\bigg(\frac{Q^2}{M\theta}\bigg)\bigg\}-\frac{3Q^{2}G_{5}}{\pi^{2}R^{6}}\bigg(1-\frac{6\sigma\theta}{R}\bigg)\bigg[1-\bigg\{1+\bigg(1+\frac{2\sigma\theta}{R}\bigg)\frac{R^2}{4\theta}\bigg\}e^{-\frac{R^2}{4\theta}}\bigg]^2\Bigg]\bigg(\frac{\rho }{\lambda }\bigg)^\frac{1}{3}. \nonumber \\
\label{dp39}
\end{eqnarray}
This is the sought relation between the critical temperature and the charge density. Note that we have used $r_{+(c)}\approx R$ (which is the leading order term in eq.(\ref{dp37})) in the exponential term $e^{-\frac{r_{+(c)}^2}{4\theta}}$.\\
\noindent In the subsequent discussion, we set $m^{2}=-3$. This particular choice of $m^2$ gives $\Delta_{+}=3$ from eq.(\ref{d12}). Eq.(\ref{dp30}) therefore takes the form
\begin{eqnarray}
p(z)=z^{3}\Bigg[1&-&\Bigg\{\bigg(1+\frac{6Q^{2}G_{5}}{\pi^{2}R^{6}}F(R)\bigg)\bigg(1+e^{-\frac{R^2}{4\theta}}+\frac{R^2}{4\theta}e^{-\frac{R^2}{4\theta}}\bigg)+\frac{3Q^{2}G_{5}}{4\theta\pi^{2}R^{4}}\bigg(1-e^{-\frac{R^2}{2\theta z^2}}\bigg)\Bigg\}z^4\nonumber\\
&+&\frac{6Q^{2}G_{5}}{\pi^{2}R^{6}}z^{6}\bigg(1-e^{-\frac{R^2}{4\theta z^2}}\bigg)^{2}\Bigg]\nonumber
\end{eqnarray}
\begin{eqnarray}
q(z)=-3z\Bigg[3z^{4}\Bigg\{\bigg(1&+&\frac{6Q^{2}G_{5}}{\pi^{2}R^{6}}F(R)\bigg)\bigg(1+e^{-\frac{R^2}{4\theta}}+\frac{R^2}{4\theta}e^{-\frac{R^2}{4\theta}}\bigg)+\frac{3Q^{2}G_{5}}{4\theta\pi^{2}R^{4}}\bigg(1-e^{-\frac{R^2}{2\theta z^2}}\bigg)\Bigg\}\nonumber\\
&-&\frac{30Q^{2}G_{5}}{\pi^{2}R^{6}}z^{6}\bigg(1-e^{-\frac{R^2}{4\theta z^2}}\bigg)^{2}\Bigg]\nonumber
\end{eqnarray}
\begin{eqnarray}
r(z)=\frac{z^{3}\bigg\{(1-z^{2})^{2} - b(\lambda^2|_{b=0}) (1-z^{2}) (1-z^{8})\bigg\}}{1-z^{4}\Bigg\{\bigg(1+\frac{6Q^{2}G_{5}}{\pi^{2}R^{6}}F(R)\bigg)\bigg(1+e^{-\frac{R^2}{4\theta}}+\frac{R^2}{4\theta}e^{-\frac{R^2}{4\theta}}\bigg)+\frac{3Q^{2}G_{5}}{4\theta\pi^{2}R^{4}}\bigg(1-e^{-\frac{R^2}{2\theta z^2}}\bigg)\Bigg\}
+\frac{6Q^{2}G_{5}}{\pi^{2}R^{6}}z^{6}\bigg(1-e^{-\frac{R^2}{4\theta z^2}}\bigg)^{2}}.\nonumber \\
\label{dp40}
\end{eqnarray}
Using these functions, we can now calculate the value of $\lambda^2$ from eq.(\ref{dp31}) by following the procedure in \cite{sgdc1}. This gives us the relation between the critical temperature and the charge density for different values of $b$, $Q$ and $\theta$. It is reassuring to note that for $Q \to 0$, we recover the results in \cite{dg2}. Our analytical results and numerical results are shown in Table \ref{tab1} and Table \ref{tab1a}. We have displayed our findings in Figure \ref{dpfig1} where we have plotted our analytical values of $\frac{T_c}{\rho^{1/3}}$ vs. $Q$ for different values of the BI parameter $b$, NC parameter $\theta$ and the mass $M$ of the black hole. We observe that the effect of charge on the critical temperature is effective when the mass of black hole is small. It is also observed that the value of $\frac{T_{c}}{\rho^{\frac{1}{3}}}$ decreases for increasing values of the charge of the black hole $Q$ and the parameters $b$ and $\theta$. 

\begin{table}[h]
\caption{Analytical results for the critical temperature and the charge density $\left(\frac{T_{c}}{\rho^{\frac{1}{3}}}\right)$ for different values of $M$, $Q$, $\theta$ and $b$}
\centering                       
\begin{tabular}{|c| c| c| c| c| c| c| c| c| c| c|}
\hline
$\frac{Q}{\sqrt{G_{5}}}$ & $\theta$ & \multicolumn{3}{c|}{$M=\frac{50}{G_{5}}$} & \multicolumn{3}{c|}{$M=\frac{100}{G_{5}}$} & \multicolumn{3}{c|}{$M=\frac{150}{G_{5}}$} \\
\hhline{~~---------}
  &  & {b=0} & {b=0.01} & {b=0.02}& {b=0} & {b=0.01} & {b=0.02}& {b=0} & {b=0.01} & {b=0.02}\\
\hline 
0.0 & 0.3 & 0.1946 & 0.1835 & 0.1680 & 0.1961 & 0.1849 & 0.1693 & 0.1962 & 0.1850 & 0.1693 \\
    & 0.5 & 0.1798 & 0.1696 & 0.1554 & 0.1920 & 0.1811 & 0.1658 & 0.1950 & 0.1838 & 0.1682 \\
    & 0.7 & 0.1621 & 0.1530 & 0.1407 & 0.1803 & 0.1701 & 0.1558 & 0.1885 & 0.1777 & 0.1627 \\
    & 0.9 & 0.1501 & 0.1423 & 0.1318 & 0.1667 & 0.1575 & 0.1446 & 0.1779 & 0.1679 & 0.1538  \\
\hline
10  & 0.3 & 0.1885 & 0.1779 & 0.1632 & 0.1941 & 0.1831 & 0.1677 & 0.1951 & 0.1840 & 0.1685  \\ 
    & 0.5 & 0.1740 & 0.1641 & 0.1506 & 0.1900 & 0.1793 & 0.1642 & 0.1938 & 0.1828 & 0.1674 \\
    & 0.7 & 0.1559 & 0.1474 & 0.1358 & 0.1783 & 0.1683 & 0.1542 & 0.1874 & 0.1768 & 0.1619 \\
    & 0.9 & 0.1453 & 0.1380 & 0.1280 & 0.1650 & 0.1557 & 0.1431 & 0.1768 & 0.1668 & 0.1529 \\
\hline 
20  & 0.3 & 0.1596 & 0.1513 & 0.1399 & 0.1871 & 0.1766 & 0.1621 & 0.1916 & 0.1808 & 0.1657 \\
    & 0.5 & 0.1484 & 0.1405 & 0.1300 & 0.1832 & 0.1730 & 0.1588 & 0.1904 & 0.1796 & 0.1646 \\
    & 0.7 & 0.1311 & 0.1244 & 0.1154 & 0.1716 & 0.1621 & 0.1490 & 0.1839 & 0.1736 & 0.1592  \\
    & 0.9 & 0.1213 & 0.1157 & 0.1082 & 0.1580 & 0.1494 & 0.1375 & 0.1734 & 0.1637 & 0.1502 \\
\hline
\end{tabular}
\label{tab1}
\end{table}

\begin{table}[h]
\caption{Numerical results for the critical temperature and the charge density $\left(\frac{T_{c}}{\rho^{\frac{1}{3}}}\right)$ for different values of $M$, $Q$, $\theta$ and $b$}
\centering                       
\begin{tabular}{|c| c| c| c| c| c| c| c| c| c| c|}
\hline
$\frac{Q}{\sqrt{G_{5}}}$ & $\theta$ & \multicolumn{3}{c|}{$M=\frac{50}{G_{5}}$} & \multicolumn{3}{c|}{$M=\frac{100}{G_{5}}$} & \multicolumn{3}{c|}{$M=\frac{150}{G_{5}}$} \\
\hhline{~~---------}
  &  & {b=0} & {b=0.01} & {b=0.02}& {b=0} & {b=0.01} & {b=0.02}& {b=0} & {b=0.01} & {b=0.02}\\
\hline 
0.0 & 0.3 & 0.1946 & 0.1835 & 0.1680 & 0.1961 & 0.1849 & 0.1693 & 0.1962 & 0.1850 & 0.1694 \\
    & 0.5 & 0.1802 & 0.1701 & 0.1560 & 0.1921 & 0.1812 & 0.1659 & 0.1957 & 0.1846 & 0.1690 \\
    & 0.7 & 0.1627 & 0.1539 & 0.1418 & 0.1807 & 0.1706 & 0.1564 & 0.1923 & 0.1814 & 0.1661 \\
    & 0.9 & 0.1522 & 0.1444 & 0.1339 & 0.1677 & 0.1585 & 0.1458 & 0.1848 & 0.1744 & 0.1598  \\
\hline
10  & 0.3 & 0.1885 & 0.1779 & 0.1632 & 0.1941 & 0.1831 & 0.1677 & 0.1955 & 0.1844 & 0.1688  \\ 
    & 0.5 & 0.1743 & 0.1647 & 0.1513 & 0.1901 & 0.1794 & 0.1643 & 0.1950 & 0.1840 & 0.1684 \\
    & 0.7 & 0.1568 & 0.1485 & 0.1371 & 0.1787 & 0.1687 & 0.1548 & 0.1916 & 0.1808 & 0.1655 \\
    & 0.9 & 0.1466 & 0.1391 & 0.1293 & 0.1657 & 0.1567 & 0.1442 & 0.1842 & 0.1738 & 0.1593 \\
\hline 
20  & 0.3 & 0.1596 & 0.1513 & 0.1399 & 0.1871 & 0.1766 & 0.1621 & 0.1933 & 0.1824 & 0.1671 \\
    & 0.5 & 0.1484 & 0.1405 & 0.1300 & 0.1832 & 0.1730 & 0.1588 & 0.1904 & 0.1796 & 0.1646 \\
    & 0.7 & 0.1311 & 0.1244 & 0.1154 & 0.1716 & 0.1621 & 0.1490 & 0.1839 & 0.1736 & 0.1592  \\
    & 0.9 & 0.1213 & 0.1157 & 0.1082 & 0.1580 & 0.1494 & 0.1375 & 0.1734 & 0.1637 & 0.1502 \\
\hline
\end{tabular}
\label{tab1a}
\end{table}

\section{Condensation values and the critical exponent}
\noindent In this section, we shall investigate the relation between condensation operator and the critical temperature. To proceed, we write down the field equation (\ref{dp13}) for $\phi (z)$  near the critical temperature $T_{c}$  
\begin{eqnarray}
\phi^{\prime \prime}(z) - \frac{d-4}{z} \phi^{\prime}(z) + \frac{d-2}{r^2_{+}} b \phi^{\prime}(z)^{3} z^3 = \frac{\langle J\rangle ^{2}}{r^{2}_{+}} \mathcal{B}(z)\phi (z)
\label{d16} 
\end{eqnarray}
where
\begin{eqnarray}
\mathcal{B}(z)= \frac{2 z^{2\Delta_{+} -4}}{r_{+}^{2\Delta_{+} -4}}\frac{F^{2}(z)}{f(z)}\left( 1- \frac{b z^4}{r^2_{+}}\phi^{\prime}(z)^2 \right)^{\frac{3}{2}}.
\end{eqnarray}  
Note that we have kept the general form for the black hole spacetime ($f(z)$) which would be later set as the NC inspired RN metric. The next step is to expand $\phi(z)$ in the small parameter $\frac{\langle J \rangle ^2}{r^{2}_{+}}$ as 
\begin{eqnarray}
\frac{\phi(z)}{r_{+}} = \lambda \left\{ (1-z^{d-3}) - \frac{b(\lambda^2|_{b=0}) (d-3)^3}{2(3d-7)} (1-z^{3d-7})\right\} + \frac{\langle J\rangle ^2}{r^{2}_{+}} \zeta (z)
\label{d15} 
\end{eqnarray}
with $\zeta (1)= 0 =\zeta^{\prime}(1).$\\
Substituting eq.(\ref{d15}) in eq.(\ref{d16}) and comparing the coefficient of $\frac{\langle J\rangle ^2}{r^{2}_{+}}$ on both sides of this equation (keeping terms upto $\mathcal{O}(b)$), we get the equation for the correction $\zeta(z)$ near the critical temperature
\begin{eqnarray}
\zeta^{\prime \prime}(z) -\left\{ \frac{d-4}{z} + 3b(\lambda^2|_{b=0}) (d-2)(d-3)^2 z^{2d-5} \right\} \zeta^{\prime}(z) = \lambda \frac{2 z^{2\Delta_{+} -4}}{r_{+}^{2\Delta_{+} -4}}\frac{F^{2}(z)}{f(z)}\mathcal{A}_{1} (z)
\label{cx3}
\end{eqnarray}
where
\begin{eqnarray}
\mathcal{A}_{1} (z) = 1-z^{d-3} -\frac{3b(\lambda^2|_{b=0}) (d-3)^2}{2}\left\{(1-z^{d-3})z^{2d-4} +\frac{d-3}{3(3d-7)}(1-z^{3d-7}) \right\}.
\end{eqnarray}
To solve this equation, we multiply it by the integrating factor $z^{-(d-4)} e^{\frac{3(d-2)(d-3)^2 b(\lambda^2|_{b=0})}{2d-4} z^{2d-4}} $ to get
\begin{eqnarray}
\frac{d}{dz}\left( z^{-(d-4)} e^{\frac{3(d-2)(d-3)^2 b(\lambda^2|_{b=0})}{2d-4} z^{2d-4}} \zeta^{\prime}(z) \right) = \lambda \frac{2 z^{2\Delta_{+} -2}}{r_{+}^{2\Delta_{+} -2}}\frac{z^{-(d-4)} F^{2}(z)}{g_{0}(z)}  e^{\frac{3(d-2)(d-3)^2 b(\lambda^2|_{b=0}) z^{2d-4}}{2d-4} } \mathcal{A}_{1} (z). 
\label{cx4}
\end{eqnarray}
Using the boundary conditions on $\zeta(z)$, we integrate the above equation between the limits $z=0$ and $z=1$. This gives
\begin{eqnarray}
\frac{\zeta^{\prime}(z)}{z^{d-4}}\mid_{z\rightarrow 0} = -\frac{\lambda}{r^{2\Delta_{+} -2}_{+}} \mathcal{A}_{2}
\label{cx5}
\end{eqnarray}
where 
\begin{eqnarray}
\mathcal{A}_{2} = \int^{1}_{0} dz \frac{2 z^{2\Delta_{+}-2}z^{-(d-4)} F^{2}(z)}{g_{0}(z)}  e^{\frac{3(d-2)(d-3)^2 b(\lambda^2|_{b=0})}{2d-4} z^{2d-4}} \mathcal{A}_{1}(z).
\end{eqnarray}
\noindent The asymptotic behaviour of $\phi(z)$ in eq.(\ref{dp10}) and eq.(\ref{d15}) gives the following equation
\begin{eqnarray}
\mu -\frac{\rho}{r^{d-3}_{+}}z^{d-3} &=& \lambda r_{+} \left\{ (1-z^{d-3}) - \frac{b(\lambda^2|_{b=0}) (d-3)^3}{2(3d-7)} (1-z^{3d-7})\right\} \nonumber \\
&+& \frac{\langle J\rangle ^2}{r_{+}} \left\{\zeta(0)+z\zeta^{\prime}(0)+......+\frac{\zeta^{d-3}(0)}{(d-3)!} z^{d-3}+....\right\}.
\label{cx7}
\end{eqnarray}
Comparing the coefficient of $z^{d-3}$ on both sides of this equation, we get
\begin{eqnarray}
\label{cx8}
-\frac{\rho}{r^{d-3}_{+}} = -\lambda r_{+} + \frac{\langle J\rangle ^2}{r_{+}}\frac{\zeta^{d-3}(0)}{(d-3)!}~.\\
\zeta^{\prime}(0)=\zeta^{\prime\prime}(0)=.......\zeta^{d-4}(0)=0.
\end{eqnarray} 
It is to be noted that $\zeta^{\prime}(z)$ and $(d-3)$-th derivative of $\zeta(z)$ are related by
\begin{eqnarray}
\frac{\zeta^{(d-3)}(z=0)}{(d-4)!} = \frac{\zeta^{\prime}(z)}{z^{d-4}}|_{z\rightarrow 0}.
\label{cx6}
\end{eqnarray}
Eq.(s)(\ref{cx5}, \ref{cx8},\ref{cx6}) yields the relation between the charge density $\rho$ and the condensation operator $\langle J\rangle$ 
\begin{eqnarray}
\frac{\rho}{r^{d-2}_{+}} = \lambda\left[1 + \frac{\langle J\rangle ^2}{r^{2\Delta_{+}}_{+}}\frac{\mathcal{A}_{2}}{(d-3)} \right].
\label{cx9}
\end{eqnarray}  
Simplifying the above relation and using eq.(\ref{dp6}) and eq.(\ref{dp20}), we obtain
\begin{eqnarray}
\langle J\rangle ^2 &=& \frac{(d-3)(4\pi T_{c})^{2\Delta_{+}}}{\mathcal{A}_{2}\Bigg[(d-1)-\frac{4MG_{d}}{\Gamma (\frac{d-1}{2})}\frac{e^{-{\frac{r_{+(c)}^2}{4\theta}}}}{(4\theta)^{\frac{d-1}{2}}}\bigg\{1-\frac{(d-2)(d-3)\Gamma (\frac{d-3}{2})}{2^{\frac{3d-7}{2}}\pi^{d-3}\theta^{\frac{d-3}{2}}}\frac{Q^2}{M}\bigg\}-\frac{2(d-2)(d-3)Q^{2}G_{d}}{\pi^{d-3}r_{+(c)}^{2(d-2)}}\gamma ^{2}\bigg(\frac{d-1}{2},\frac{r_{+(c)}^2}{4\theta}\bigg)\Bigg]^{2\Delta_{+}}}\nonumber\\
&&~~~~~~~~~~\times\left(\frac{T_{c}}{T}\right)^{d-2} \left[1- \left(\frac{T}{T_{c}}\right)^{d-2} \right].
\label{cx10}
\end{eqnarray}
Since $T \approx T_{c}$, we have 
\begin{eqnarray}
\left(\frac{T_{c}}{T}\right)^{d-2} \left[1- \left(\frac{T}{T_{c}}\right)^{d-2} \right] \approx  (d-2)\left[1- \left(\frac{T}{T_{c}}\right)\right].
\label{cx11}
\end{eqnarray} 
From this we finally obtain the relation between the condensation operator and the critical temperature in $d$-dimensions
\begin{eqnarray}
\langle J\rangle = \beta T^{\Delta_{+}}_{c} \sqrt{1-\frac{T}{T_{c}}}
\label{cx12}
\end{eqnarray}
where
\begin{eqnarray}
\beta = \sqrt{\frac{(d-3)(d-2)}{\mathcal{A}_{2}}} \left[\frac{4\pi}{(d-1)-\frac{4MG_{d}}{\Gamma (\frac{d-1}{2})}\frac{e^{-{\frac{r_{+(c)}^2}{4\theta}}}}{(4\theta)^{\frac{d-1}{2}}}\bigg\{1-\frac{(d-2)(d-3)\Gamma (\frac{d-3}{2})}{2^{\frac{3d-7}{2}}\pi^{d-3}\theta^{\frac{d-3}{2}}}\frac{Q^2}{M}\bigg\}-\frac{2(d-2)(d-3)Q^{2}G_{d}}{\pi^{d-3}r_{+(c)}^{2(d-2)}}\gamma ^{2}\bigg(\frac{d-1}{2},\frac{r_{+(c)}^2}{4\theta}\bigg)}\right]^{\Delta_{+}}.
\end{eqnarray} 
The critical exponent is observed to be $1/2$
which agrees with the universal mean field value.
We shall now set $d=5$ and $m^2=-3$ for the rest of our discussion. The choice for $m^2$ yields $\Delta_{+}=3$. Eq.(\ref{cx12}) now simplifies to 
\begin{eqnarray}
\langle J\rangle = \beta T^{3}_{c} \sqrt{1-\frac{T}{T_{c}}}~.
\label{cx13}
\end{eqnarray}
The expressions for $\mathcal{A}_{1}(z)$ and $\beta$ simplify to 
\begin{eqnarray}
\label{dp54}
\mathcal{A}_{1} (z)
&=& (1-z^2)\left[1-\frac{b(\lambda^2|_{b=0})}{2} (1+z^2 +z^4 +13z^6) \right]  \\
\beta &=& \sqrt{\frac{6}{\mathcal{A}_{2}}} \left[\frac{\pi}{1-\frac{MG_{5}}{(4\theta)^2}e^{-\frac{R^2}{4\theta}}\bigg\{1-\frac{3}{8\pi^2}\bigg(\frac{Q^2}{M\theta}\bigg)\bigg\}-\frac{3Q^{2}G_{5}}{\pi^{2}R^{6}}\bigg(1-\frac{6\sigma\theta}{R}\bigg)\bigg[1-\bigg\{1+\bigg(1+\frac{2\sigma\theta}{R}\bigg)\frac{R^2}{4\theta}\bigg\}e^{-\frac{R^2}{4\theta}}\bigg]^2} \right]^{3} \nonumber \\
&=& \beta_{1} \pi^3
\end{eqnarray}
Substituting eq.(\ref{dp54}) in $\mathcal{A}_{2}$ and keeping terms upto $\mathcal{O}(b)$, we obtain 
\begin{eqnarray}
\mathcal{A}_{2} &=& \int^{1}_{0} dz \frac{2 z^{3} F^{2}(z)}{g_{0}(z)}  e^{6 b(\lambda^2|_{b=0}) z^{6}} \mathcal{A}_{1}(z) \nonumber\\
 &\approx& \int^{1}_{0} dz \frac{2 z^{3} F^{2}(z) (1-z^2)}{g_{0}(z)} \left\{1- \frac{b(\lambda^2|_{b=0})}{2} (1+z^2 +z^4 +z^6) \right\}
\end{eqnarray}
For $Q=0$ we recover the results in \cite{dg2}. In Tables \ref{dptab2} 
and \ref{dptab4}, we have shown the numerical and analytical values of $\beta_{1}$ for different values $b$, $\theta$, $M$ and $Q$ and we find that they are in very good agreement.

\begin{table}[h!]
\caption{Numerical results for condensation operator values $\beta_{1}$ for different values of $M$, $Q$, $\theta$ and $b$}
\centering                          
\begin{tabular}{|c| c| c| c| c| c| c| c| c| c| c| }
\hline
$\frac{Q}{\sqrt{G_{5}}}$& $\theta$ & \multicolumn{3}{c|}{$M=\frac{50}{G_{5}}$} & \multicolumn{3}{c|}{$M=\frac{100}{G_{5}}$} & \multicolumn{3}{c|}{$M=\frac{150}{G_{5}}$} \\
\hhline{~~---------}
  &  & {b=0} & {b=0.01} & {b=0.02}& {b=0} & {b=0.01} & {b=0.02}& {b=0} & {b=0.01} & {b=0.02}\\
\hline 
0.0 & 0.3 &  7.893 &  8.949 & 10.615 &  7.717 &  8.748 & 10.375 &  7.705 &  8.735 & 10.359 \\
    & 0.5 &  9.856 & 11.191 & 13.295 &  8.192 &  9.29 & 11.022 &  7.756 &  8.793 & 10.428 \\
    & 0.7 & 13.145 & 14.958 & 17.808  & 9.782 & 11.106 & 13.193 &  8.166 &  9.261 & 10.987 \\
    & 0.9 & 15.578 & 17.760 & 21.172 & 12.092 & 13.75 & 16.360 & 9.163 & 10.4 & 12.348 \\
\hline
10  & 0.3 &  8.650 &  9.805 & 11.621 &  7.952 &  9.015 & 10.688 &  7.785 &  8.826 & 10.466 \\ 
    & 0.5 & 10.848 & 12.315 & 14.621 &  8.445 &  9.576 & 11.359 &  7.837 &  8.884 & 10.535 \\
    & 0.7 & 14.582 & 16.592 & 19.743 & 10.1 & 11.465 & 13.617 &  8.253 & 9.359 & 11.102 \\
    & 0.9 & 17.335 & 19.763 & 23.546 & 12.507 & 14.222 & 16.918 & 9.263 & 10.512 & 12.482 \\
\hline 
20  & 0.3 & 13.935 & 15.792 & 18.668 &  8.844 & 10.024 & 11.874 &  8.046 &  9.120 & 10.811 \\
    & 0.5 & 17.077 & 19.381 & 22.952 &  9.386 & 10.642 & 12.612 &  8.1 &  9.180 & 10.883 \\
    & 0.7 & 24.128 & 27.451 & 32.592 & 11.263 & 12.785 & 15.174 &  8.531 & 9.673 & 11.473 \\
    & 0.9 & 29.593 & 33.717 & 40.06 & 14.042 & 15.965 & 18.98 & 9.584 & 10.876 & 12.91 \\
\hline
\end{tabular}
\label{dptab2}
\end{table}
\begin{table}[h!]
\caption{Analytical results for condensation operator values $\beta_{1}$ for different values of $M$, $Q$, $\theta$ and $b$}
\centering                          
\begin{tabular}{|c| c| c| c| c| c| c| c| c| c| c| }
\hline
$\frac{Q}{\sqrt{G_{5}}}$& $\theta$ & \multicolumn{3}{c|}{$M=\frac{50}{G_{5}}$} & \multicolumn{3}{c|}{$M=\frac{100}{G_{5}}$} & \multicolumn{3}{c|}{$M=\frac{150}{G_{5}}$} \\
\hhline{~~---------}
  &  & {b=0} & {b=0.01} & {b=0.02}& {b=0} & {b=0.01} & {b=0.02}& {b=0} & {b=0.01} & {b=0.02}\\
\hline 
0.0 & 0.3 &  7.893 &  8.955 & 10.627 &  7.717 &  8.748 & 10.375 &  7.706 &  8.736 & 10.361 \\
    & 0.5 &  9.819 & 11.276 & 13.571 &  8.221 &  9.326 & 11.068 &  7.866 &  8.916 & 10.570 \\
    & 0.7 & 13.460 & 15.693 & 19.13  & 10.052 & 11.403 & 13.527 &  8.723 &  9.894 & 11.737 \\
    & 0.9 & 21.755 & 23.407 & 25.629 & 13.022 & 14.744 & 17.421 & 10.066 & 11.605 & 14.029 \\
\hline
10  & 0.3 &  8.657 &  9.816 & 11.640 &  7.951 &  9.014 & 10.688 &  7.831 &  8.877 & 10.526 \\ 
    & 0.5 & 11.290 & 12.760 & 15.059 &  8.472 &  9.612 & 11.406 &  7.987 &  9.055 & 10.739 \\
    & 0.7 & 16.230 & 18.340 & 21.549 & 10.218 & 11.662 & 13.930 &  8.812 & 10.018 & 11.918 \\
    & 0.9 & 22.017 & 24.366 & 27.830 & 13.105 & 15.004 & 17.954 & 10.435 & 11.946 & 14.319 \\
\hline 
20  & 0.3 & 13.936 & 15.791 & 18.667 &  8.848 & 10.026 & 11.875 &  8.254 &  9.355 & 11.088 \\
    & 0.5 & 17.244 & 19.734 & 23.595 &  9.350 & 10.633 & 12.648 &  8.412 &  9.538 & 11.310 \\
    & 0.7 & 26.807 & 30.442 & 35.943 & 11.545 & 13.118 & 15.580 &  9.638 & 10.812 & 12.656 \\
    & 0.9 & 42.067 & 45.176 & 49.310 & 16.521 & 18.137 & 20.615 & 11.276 & 12.805 & 15.200 \\
\hline
\end{tabular}
\label{dptab4}
\end{table}


\section{Conclusions}
In this paper, we have analytically investigated the effects of the charge and mass of a black hole on holographic superconductors in the presence of a noncommutative inspired Reissner-Nordstr\"{o}m black hole. Using the Sturm-Liouville eigenvalue method, the relation between the critical temperature and charge density is obtained first. It is observed that the condensation gets hard to form in the presence of the Born-Infeld ($b$) and noncommutative ($\theta$) parameters. Further, it is found that higher values of the charge and mass of the black hole makes the condensate even harder to form. However, for large mass black holes, the effect of charge of the black hole is negligible as seen from the relation between the critical temperature and the charge density. We also conclude from our investigations that the charge of the black hole plays a crucial role on the properties of holographic superconductors when the mass of the black hole is small. From the expression of the condensation operator, we observe that the values of the condensation operator are higher for higher values of $b, ~\theta, ~M$ and $Q$. Our analytical results agree very well with the numerical results.

\section*{Acknowledgments}
DP would like to thank CSIR for financial support. DG would like to thank DST-INSPIRE for financial support. S.G. acknowledges the support by DST SERB under Start Up Research Grant (Young Scientist), File No.YSS/2014/000180. SG also acknowledges the support under the Visiting Associateship programme of IUCAA, Pune.


\end{document}